\documentclass[peerreview,conference,a4paper]{IEEEtran} 
\usepackage{gensymb}
\usepackage{comment}
\usepackage{color}
\usepackage[table]{xcolor}
\usepackage{subfigure}
\usepackage{amsmath}
\usepackage{graphicx}
\usepackage{epstopdf}
\usepackage{verbatim}
\usepackage[ansinew]{inputenc}
\usepackage{epsfig}
\usepackage{amsfonts}
\usepackage{amssymb}
\usepackage{setspace}
\usepackage{soul}
\usepackage{multirow}
\usepackage{wrapfig}
\usepackage{algpseudocode}
\usepackage[lined,boxed,commentsnumbered,ruled]{algorithm2e}
\usepackage{cite} 
\usepackage{url}  

\usepackage{ifthen}  
\usepackage{multicol}   
\usepackage[]{inputenc} 
\usepackage{authblk}
\usepackage{caption}
\IEEEoverridecommandlockouts

\begin{document}
%
\title{Crowdsourcing On-street Parking Space Detection}


\author[1]{Ruizhi Liao}
\author[1]{Cristian Roman}
\author[1]{Peter Ball}
\author[1]{Shumao Ou}
\author[2]{Liping Chen}
\affil[1]{Department of Computing and Communication Technologies, Oxford Brookes University, United Kingdom}
\affil[2]{Industry Solution, IBM Analytics, Ireland}

\affil[ ]{\textit {\{rliao,pball,sou\}@brookes.ac.uk}, cristian@cristianroman.com, lipingch@ie.ibm.com}


\maketitle

\begin{abstract}

As the number of vehicles continues to grow, parking spaces are at a premium in city streets. Additionally, due to the lack of knowledge about street parking spaces, heuristic circling the blocks not only costs drivers' time and fuel, but also increases city congestion. In the wake of recent trend to build convenient, green and energy-efficient smart cities, we rethink common techniques adopted by high-profile smart parking systems, and present a user-engaged (crowdsourcing) and sonar-based prototype to identify urban on-street parking spaces. The prototype includes an ultrasonic sensor, a GPS receiver and associated Arduino micro-controllers. It is mounted on the passenger side of a car to measure the distance from the vehicle to the nearest roadside obstacle. Multiple road tests are conducted around Wheatley, Oxford to gather results and  emulate the crowdsourcing approach. By extracting parked vehicles' features from the collected trace, a supervised learning algorithm is developed to estimate roadside parking occupancy and spot illegal parking vehicles. A quantity estimation model is derived to calculate the required number of sensing units to cover urban streets. The estimation is quantitatively compared to a fixed sensing solution. The results show that the crowdsourcing way would need substantially fewer sensors compared to the fixed sensing system. 

\end{abstract}

\begin{keywords}
on-street/roadside parking, crowdsourcing, sonar/ultrasonic, supervised learning
\end{keywords}

\IEEEpeerreviewmaketitle


\section{Introduction}\label{sec:intro}

Due to the lack of real-time knowledge about the roadside parking occupancy, circling around urban streets to find a parking space not only wastes time and fuel, but also increases traffic flows. The continuous growth of vehicles and the waves of urbanization made the parking problem worse. A downtown traffic study \cite{shoup2006cruising} on several major cities reveals that cruising for kerb vacancies is an often-overlooked source of congestion, which accounts for up to $30$ percent of total traffic flows\footnote{The surveyed cities include New York, San Francisco, Freiburg, Jerusalem, Cape Town, Sydney and London.}. In order to quantify the wasted time and fuel, we made an intuitive calculation  based on the JustPark\footnote{www.justpark.com.} estimation that heuristic searching takes an average of $6.75$ minutes to find a parking spot in the UK. Assuming the turnover of a parking spot is $10$ cars a day, cruising for parking costs $67.5$ minutes per spot a day ($24637.5$ minutes a year). If the cruising speed is $15$ kilometers per hour (km/h), it generates $16.87$ km per spot a day ($6157.55$ km a year), or $615.755$ litres fuel a year for a single parking spot - based on a consumption of $10$ litres fuel per $100$ km at the speed of $15$ km/h. Imagine the number of roadside parking spots worldwide, and its impact in both ecological and economical terms.

The frustrating circling, and the above-mentioned time or fuel wastes are exactly what an Intelligent Transportation System (ITS)  aims to address, or broadly speaking, to build convenient, green and energy-efficient Smart Cities. There are various ITS or smart city projects related to parking monitoring, such as private parking \cite{Amsterdam}\cite{eu}, off-street parking \cite{6583499}\cite{mahmud2013survey} and on-street parking \cite{sf}\cite{ws}. 

Private parking originates from peer-to-peer based sharing economy, whose concept is to rent out owners' spare facilities, e.g., Airbnb for rooms, Uber for cars, and JustPark for empty driveways. Off-street parking refers to multi-storey car parks or large fields that can accommodate hundreds of vehicles. The occupancy of off-street parks can be easily obtained by applying entrance counters, acoustic or vision based sensing techniques. The parking availability  is disseminated via APPs or web portals (e.g., Parker and Parkopedia). On-street or roadside parking refers to parking spaces along public roads. It accounts for a considerable fraction of urban parking, and moreover, it provides preferable access to drivers' destinations. However, on-street parking spaces are not usually monitored. 

The reasons for that are twofold. First, comparing to off-street parking, on-street parking has more severe environment, e.g., harsh weather or light conditions, which may negatively impact sensors' functionalities. Secondly, although Smart City implies the trend of using various sensing technologies to instrument the city, it is rather expensive and unscalable to dig underneaths or mount overhead for roadside parking due to safety, road implementation, maintenance and municipal coordination issues that may get involved.

For these reasons, we focus on on-street parking and present a user-engaged Crowdsourcing Parking monitoring system, which is named as CroPark. It employs ultrasonic sensors to measure the distance from the vehicle to roadside, and uses a supervised learning algorithm to estimate the number of available parking spaces. Only a reduced set of the sensed data, namely, the interpreted parked vehicles and empty spaces, together with Global Positioning System (GPS) coordinates, vehicle speed and timestamps, are transmitted to a central server, where a parking occupancy map is built. The parking availability information is updated as next CroPark vehicle passes by, and disseminated to traffic enforcement or users via a mobile APP or a web portal. 

The rest of the article is organized as follows. First, Section \ref{sec:relwork} reviews relevant on-street parking systems, and summarizes representative features of each work. By exploring other systems, the contributions of CroPark are also presented. Then, Section \ref{sec:proto} explains the CroPark system and the supervised learning algorithm. After that, Section \ref{sec:evaluation} analyses our drive-test data and presents obtained results. Finally, Section \ref{sec:conclusions} concludes the article and looks into future works.



\section{Related Work and Contributions}\label{sec:relwork}

In this section, we overview prominent on-street parking systems and summarize their representative features. By comparing CroPark with these parking systems, the contributions of our work are described. 

\newsavebox{\tablebox}
\newcommand{\tabincell}[2]{\begin{tabular}{@{}#1@{}}#2\end{tabular}}

\begin{table*}[t!!!!!!!]
\caption{{ \bfseries Comparisons of On-street Parking Projects}}
\begin{lrbox}{\tablebox}
 \rowcolors{0}{blue!10}{}
\renewcommand\arraystretch{1.9}
{\Large
\begin{tabular}{|l|l|l|}
\hline
{\bfseries On-street Parking Projects} & {\bfseries Sensing Technologies} & {\bfseries Remarks}\\ \hline
SFpark \cite{sf}, 2014 &\tabincell{l}{\vspace{-0.0em}magnetometer\\\vspace{-0.0em}$\texttt{>} 1$ sensor needed per spot} & \tabincell{l}{\vspace{-0.0em}a complete on/off-street solution;\\\vspace{-0.0em}digging roads required} \\ \hline

FASTPRK \cite{ws} &\tabincell{l}{\vspace{-0.0em}magnetic\\\vspace{-0.0em}$1$ sensor per spot}  &\tabincell{l}{\vspace{-0.0em}a company portfolio including sensing, analysis, open data interface and APP;\\\vspace{-0.0em}digging roads required} \\ \hline

Street Parking System \cite{zhang2013street}, 2013 &\tabincell{l}{\vspace{-0.0em}magnetic\\\vspace{-0.0em}$1$ sensor per spot}   &\tabincell{l}{\vspace{-0.0em}$82$ sensors are deployed at Shenzhen Institute of Advanced Technology;\\\vspace{-0.0em}digging roads required} \\ \hline

Smart Parking \cite{sp} &\tabincell{l}{\vspace{-0.0em}infrared\\\vspace{-0.0em}unspecified number}  &\tabincell{l}{\vspace{-0.0em}a company portfolio including sensing, guiding, payment and management;\\\vspace{-0.0em}digging roads required} \\ \hline

Smart Santander \cite{stander}, 2014 &\tabincell{l}{\vspace{-0.0em}ferromagnetic\\\vspace{-0.0em}$1$ sensor per spot} &\tabincell{l}{\vspace{-0.0em}part of a smart city project;\\\vspace{-0.0em}digging roads required} \\ \hline

Integrated Smart Parking \cite{ger}, ongoing &\tabincell{l}{\vspace{-0.0em}radar\\\vspace{-0.0em}$\texttt{<} 1$ sensor needed per spot}  &\tabincell{l}{\vspace{-0.0em}sensing not only parking but also traffic flows;\\\vspace{-0.0em}mounting overhead} \\ \hline

Parking Spotter \cite{ford}, ongoing &\tabincell{l}{\vspace{-0.0em}sonar or radar\\\vspace{-0.0em}$\texttt{<} 1$ sensor needed per spot} &\tabincell{l}{\vspace{-0.0em}crowdsource sensing;\\\vspace{-0.0em}Ford proprietary application} \\ \hline

ParkNet \cite{mathur2010parknet}, 2010 &\tabincell{l}{\vspace{-0.0em}sonar\\\vspace{-0.0em}$\texttt{<} 1$ sensor needed per spot} &\tabincell{l}{\vspace{-0.0em}crowdsource sensing;\\\vspace{-0.0em}environmental fingerprinting to reduce GPS errors} \\ \hline

ParkSense \cite{nawaz2013parksense}, 2013 &\tabincell{l}{\vspace{-0.0em}mobile phone and WiFi\\\vspace{-0.0em}$1$ mobile phone per spot} &\tabincell{l}{\vspace{-0.0em}use beacons between mobile phone and WiFi to infer parking status;\\\vspace{-0.0em}presence of both mobile phone and WiFi infrastructure required} \\ \hline

\end{tabular}}
\label{parkingprojects}
\end{lrbox}
\resizebox{1\textwidth}{!}{\usebox{\tablebox}}
\end{table*}

SFpark \cite{sf} is a U.S. federally-funded parking management project in San Francisco, which spans from $2009$ to $2014$ with a total budget of $46.2$ million dollars. It adopts a classical wireless sensor network structure, where the data flows from parking sensors and parking meters to a data warehouse via wireless mesh networks. The system gives a complete on/off-street parking solution and brings a list of benefits: easier to find parking (searching time decreases by $43\%$), reduced congestion (traffic volume decreases by $8\%$), lower parking rates ($11$ cents cheaper), fewer parking tickets ($23\%$ lower), et cetera. However, the cost of the high-profile parking sysem is rather high. The SFpark pilot deployment installed $11700$ magnetometer sensors and $300$ pole-mounted mesh nodes for $8000$ parking spaces (one or two sensors are installed in each parking space). From SFpark's finance figures, the parking sensors cost $5.7$ million dollars (approximately $480$ dollars each), which do not account for management, installation and maintenance expenditure. FASTPRK (magnetic) \cite{ws}, Street Parking System (magnetic) \cite{zhang2013street}, Smart Parking (infrared) \cite{sp} and Smart Santander (ferromagnetic) \cite{stander} are similar parking projects using different sensing techniques.

Integrated Smart Parking Solution \cite{ger} is a Siemens-led ongoing project aiming at simplifying the searching process, which launched a testing pilot in Berlin in September, 2015. Different from the above sensor-burying approaches, radar sensors are mouted on street lamps to scan a bigger area. The radars not only monitor traffic flows but also parking spaces, which can be either used for traffic control or facilitating drivers to find a parking spot. The benefits of the overhead radar approach are claimed to be as follows. First, it is not impaired by weather or light conditions. Secondly, it detects more than just parking spots. For example, it can measure vehicle speed, traffic flows and pedestrian flows. Thirdly, it is mounted on street lamps, which alleviates the infrastructure changes.
\begin{figure*}[t!!!!!!]
\begin{center}
\subfigure[GPS unit and Arduino Micro-controllers]{\includegraphics[scale=0.097,angle=-90]{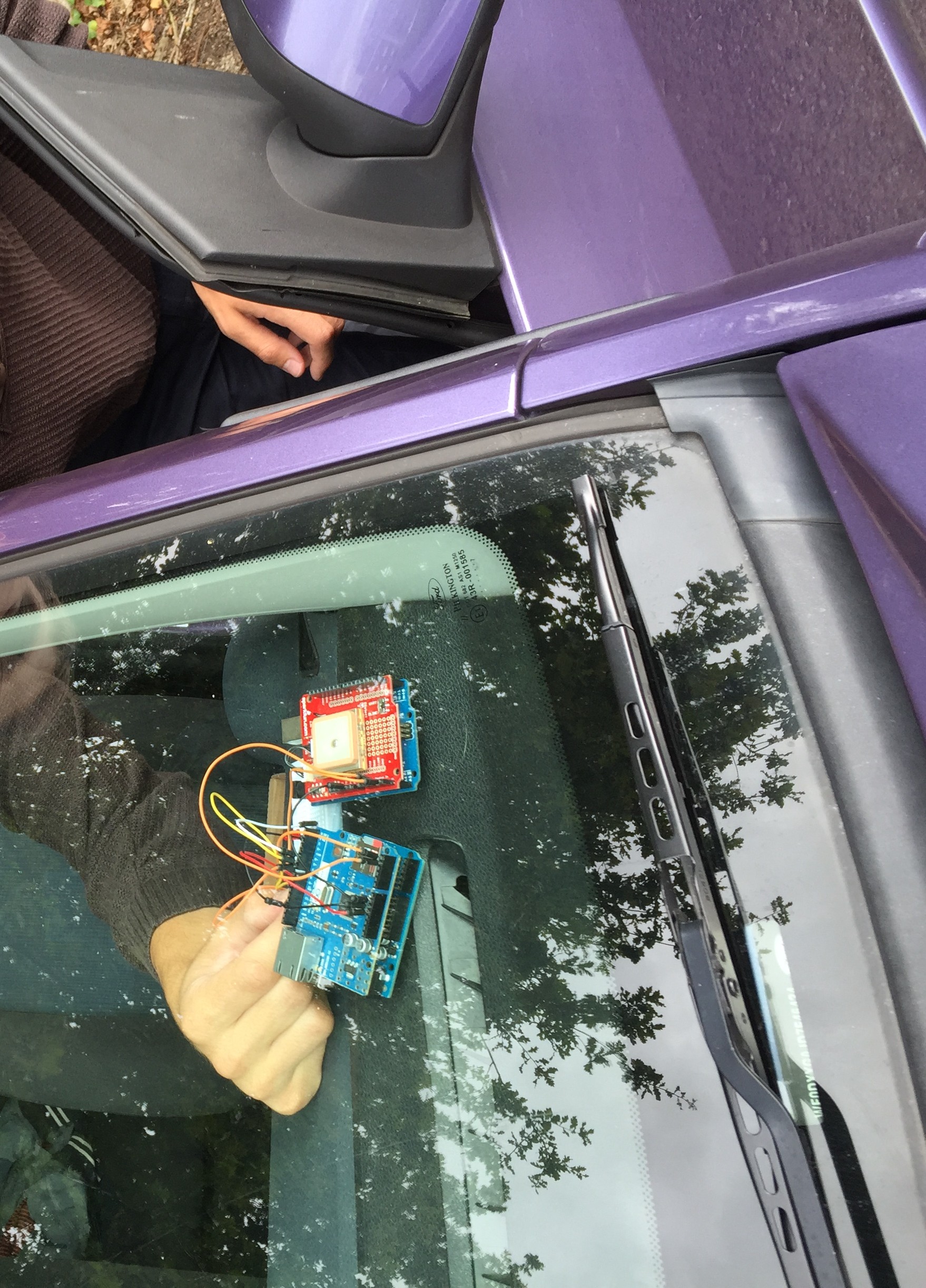}{\label{Fig:car1}}}
\subfigure[Rangefinder mounted at Side Door]{\includegraphics[scale=0.096,angle=-90]{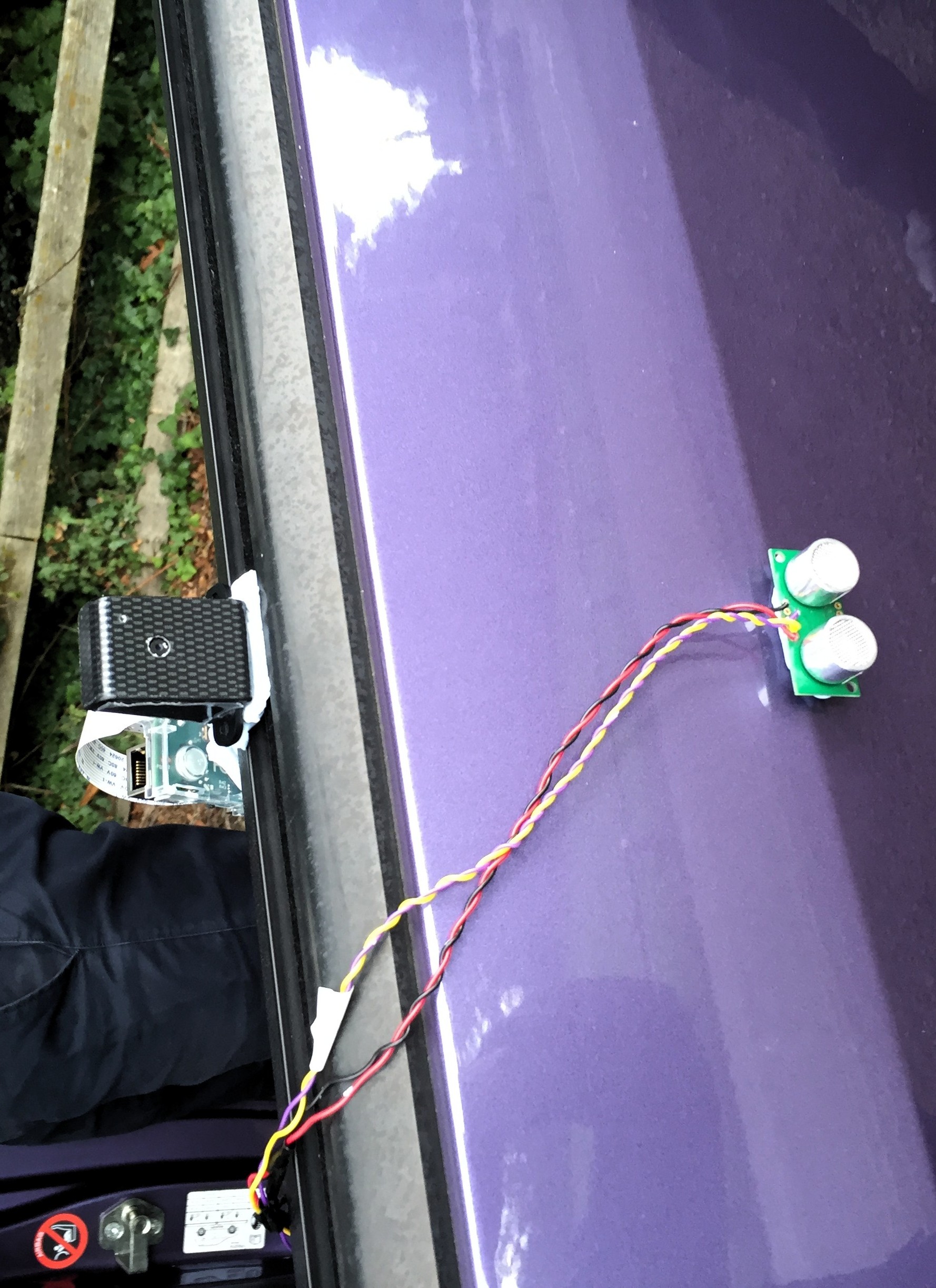}
{\label{Fig:car2}}}
\caption{Prototype Deployment}
\label{Fig:car}
\end{center}
\end{figure*}

Parking Spotter \cite{ford} is an ongoing joint work by Ford and Georgia Tech. The idea is to leverage sonars and radars that are already widely employed on Ford cars to sense the on-street parking occupancy. The sensing results are updated to a cloud server, and the results are presented to other Ford drivers as an added-value service.

Mathur et al. in \cite{mathur2010parknet} present ParkNet, a mobile street parking system, which collects parking occupancy information as vehicles pass by. ParkNet consists of a GPS receiver and an ultrasonic rangefinder. Through one-month runs of three vehicles passing by the urban streets of Highland Park, New Jersey, the authors built a parking map from collected data. In order to achieve improved location accuracy, the authors utilize an environmental fingerprinting approach, namely, using objects on the street to correct GPS errors.

Nawaz et al. in \cite{nawaz2013parksense} propose a WiFi beacon association  based sensing system named ParkSense to estimate if a driver has entered or driven away from a parking spot. More specifically, ParkSense uses the WiFi association and de-association changing rates to sense the parking status. From the empirical evaluation, the authors claim that the WiFi beacon rate is highly correlated with driver's activity. The disadvantages of the system are that 1) the presence of both mobile phone and WiFi infrastructure are required, and 2) the access to spatially distributed WiFi access points for analysis is assumed.

The features of the surveyed parking systems are summarized in Table \ref{parkingprojects}. 

The contributions of the article are summarized as follows. Some high-profile parking projects are overviewed. By looking into their features, we rethink the common sensing solutions, and present a  crowdsourcing mobile sensing technique. A supervised learning algorithm is developed for CroPark to recognize parked cars and empty spaces. The algorithm uses parked cars' contouring vector extracted from the driving  trace to train a classifier, which is used to differentiate parked cars from road clutters. A simple estimation model is proposed to calculate how many mobile sensing units are needed to cover urban streets with certain updating frequency. The estimation produced by the model is quantitatively compared to a fixed sensing solution. The results show that CroPark would need significantly lower number of sensing units than the fixed sensing system. 



\section{On-street Parking Space Detection: Prototype, Testing Scenario and Algorithm}\label{sec:proto}

\subsection{The Prototype System}
The prototype kit consists of an HC-SR04 ultrasonic rangefinder, a GPS receiver and associated Arduino micro-controllers. In the road test, we placed the prototype kit under the windshield (Fig.\ref{Fig:car1}) to provide the GPS receiver with a clear view from satellites, while we led the ultrasonic rangefinder to the passenger door (Fig.\ref{Fig:car2}) to measure the distance from the vehicle to roadside. Note the video camera above the ultrasonic sensor is not a part of the CroPark system. The employment the camera is to record the ground truth to compare it with the ultrasonic sensor data. 

The employed ultrasonic sensor is set to transmit a short pulse every $50$ milliseconds. The emitted wave is bounced back when there is an object in the way. By counting the elapsed time, together with the sound speed in the air, the distance to the object is measured. After interpolating the GPS coordinates, the format of the measurement and an example of the road test trace are given as follows.
\[
 \bigg\{  
 \begin{tabular}{l}
  $timestamp/distance/latitude/longitude/speed$ \\ \\
  $93777\text{ ms}/532 \text{ cm}/51.748295/\text{-}1.14049621/24.85 \text{ km/h}$
  \end{tabular}
\]

\begin{figure*}[t!!!!!!]
\begin{center}
\subfigure[Drive-test Routes]{\includegraphics[scale=0.63]{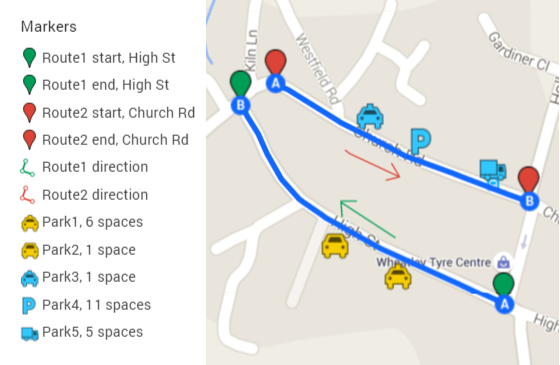}{\label{Fig:rts}}}
\subfigure[Trace of Fully-parked Route$1$]{\includegraphics[scale=0.58]{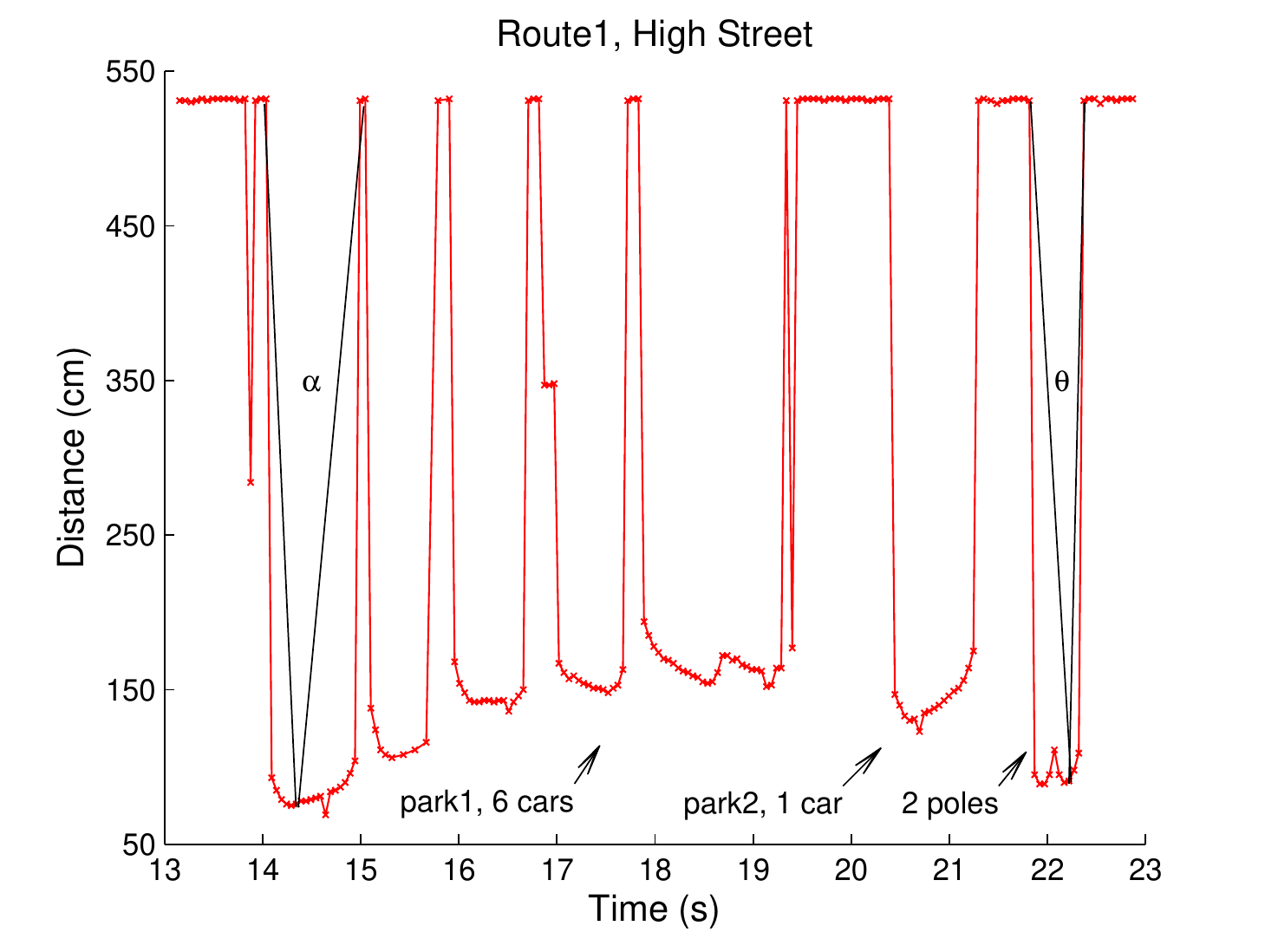}{\label{Fig:r1}}}
\caption{Drive-test Routes and Trace Plot of Route$1$}
\label{Fig:rtlot}
\end{center}
\end{figure*}

The measurement data is locally preprocessed to reduce the data traffic. In other words, only those data groups that are interpreted as parked vehicles or empty spaces by the algorithm are sent to the central database. It is proposed that the CroPark kits could be employed on public vehicles such as buses and taxis, and they continually gather data as they travel along their routes. The on-board communication links or a $3$G shield card can be used for the back-haul data transmission. The central database has pre-installed parking zones' information (e.g., the capacity and GPS coordinates of parking zones), which are used to identify illegal parking vehicles and empty parking spaces from the received sonar data. The inferred results are accessed via mobile APPs or web portals by users or traffic enforcement agents. %

\subsection{The Supervised Learning Algorithm}
A supervised learning algorithm is developed to recognise parked cars and empty spaces. We use contouring patterns that are extracted from the drive-test trace to train a classifier in the purpose of discriminating parked cars and spaces from road clutters. 

\subsubsection{Train the Classifier}
As shown in Fig.\ref{Fig:rts}, we conducted a serial of drive tests on two roads in Wheatley, Oxford. High Street (Route$1$) has two parks (park$1$ has $6$ spaces and park$2$ has $1$ space), and Church Road (Route$2$) has three parks (park$3$ has $1$ space, park$4$ has $11$ spaces and park$5$ has $5$ spaces). In our test, we repeatedly circled the two streets to emulate the crowdsourcing approach. 

The algorithm takes the drive-test trace as inputs and scans the data in sequence. An example is illustrated in Fig.\ref{Fig:r1} when park$1$ and park$2$ of the Route$1$ are fully parked. As can be observed from the figure, the sonar scanning presents parked vehicles to distinctive U-shaped contours as the testing car drives by. The contouring features allow us to distinguish parked vehicles from road furnitures (e.g., the two poles as shown in Fig.\ref{Fig:r1}). The classifier training patterns vector is summarized as follows.

\begin{itemize}
  \item Distance: From a driving-by car's perspective, there is a distance range that a parked vehicle should reside in. The lower bound of the range corresponds to lateral safety distance or minimum safety distance ($Distance_\text{min}=70$ cm). In addition to the minimum safety distance, the higher bound of the range is obtained by adding a car's width ($Distance_\text{max}=250$ cm).
    \item Standard deviation: The distance range filters out many road clutters that are outside $[Distance_\text{min},Distance_\text{max}]$, however, many still remain. Standard deviation is used to quantify the dispersion of a group of parked vehicle data. In particular, the standard deviation of the U-shaped bottom is relatively flat, which refers to the small-scale standard deviation ($\sigma_\text{small}=10.9$) and is used to describe the vehicles' core frame; while the standard deviation at the edge of the U-shaped bottom is much larger, which refers to the big-scale standard deviation ($\sigma_\text{big}=51.3$) and is used to describe the vehicles' edges accordingly.
  \item Length: Due to the fact of various vehicles' length, we adopt $2.1$ meters as a bottom line ($Length_\text{min}$) to categorize the trace for calculating the "small-scale" standard deviation, and $9$ meters as a upper bound ($Length_\text{max}$) for deriving the "large-scale" standard deviation.  
  \item Angle: The angle between vehicle's contouring vertices and bottom is calculated using trigonometric functions. The angle of the contour we obtained from the parked vehicle trace ranges from $80\degree$ to $130\degree$ (a function of speed and vehicle's length). We apply this contour angle range in the classifier to mask out road furnitures. For example, as shown in Fig.\ref{Fig:r1}, $\alpha=87.97\degree$ is the angle of first parked car, which falls in the angle range, while $\theta=55.69\degree$ is the angle of the two poles, which are masked out as noise.
\end{itemize}

\subsubsection{Classifier}
The above four contouring patters form a feature vector, which is used to train the classifier. The classifier interprets the fed data into groups of road clutters, parked vehicles or empty spaces according to the features' ascription. As illustrated in \textbf{Algorithm \ref{Alg:pant}}, the classifier iterates through the gathered trace, obtains the indices within the length range, and then checks whether the data are in line with the above-mentioned features. 

\begin{algorithm}[h!!!!!!!]

\footnotesize
\caption{Classifier Pseudocode}\label{Alg:pant}

\KwIn{$Length_\text{min,max}$, $Distance_\text{min,max}$, $\sigma_\text{small,big}$, $Speed$}

\KwOut{$vec_\text{parked}$, $vec_\text{empty}$}
\BlankLine
\For{$len = 1:length(Trace)$}
 {\BlankLine
 $Duration_\text{min,max}=Length_\text{min,max}/Speed$ \\
 $Range_\text{min,max}=$\\  $Trace(:,1) \in [Trace(len,1), Trace(len,1)+ Duration_\text{min,max}]$
  		\\
  		\Comment{$Range_\text{min,max}$ \textbf{consists of all data indices in} $Length_\text{min,max}$}
  		\BlankLine
  		\If{$std(Trace(Range_\text{min},2)) \texttt{<} \sigma_\text{small}\ \& \ std(Trace(Range_\text{max},2)) \texttt{>} \sigma_\text{big}$}  		
  		    {  		
			 \BlankLine
	    \If{$Distance \in [Distance_\text{min}, \ Distance_\text{max}]\  \& \ Angle \in [Angle_\text{min}, \ Angle_\text{max}]$}
		    		
  		       	 		{\BlankLine$vec_\text{parked}(end+1)=[Trace(len,3),\ Trace(len,4)]$}	
  		       				  		 	  		
  		    }  		      		
  		\Comment{$std( )$ \textbf{represents the standard deviation function}}
  		    	\BlankLine
			 \If{$Trace(Range_\text{max},1) \in [Trace(len,1), Trace(len,1)+ Duration_\text{max}]$}
  		    		{
  		    		\BlankLine
  		 	  	\If{$Distance \texttt{>} Distance_\text{max}$}  		   			 
  		        			{\BlankLine$vec_\text{empty}(end+1)=[Trace(len,3),\ Trace(len,4)]$}	
  		    		}
}

\end{algorithm}

\subsection{Presentation of Illegally-Parked Cars and Empty Spaces}
The software bundle, namely, Linux, Apache, MySQL, and PHP (LAMP) is employed to present the interpreted data by the classifier. An example of the developed web portal showing the parking status is shown in Fig.\ref{Fig:rts}. 

\subsubsection{Illegally-Parked Cars}
Presenting illegally-parked cars is straightforward. We first infer the illegally-parked cars by checking whether GPS coordinates of the detected cars are in the parking database. Then we mark those cars that are outside the permitted parking zones as illegal parking on the map. 
\subsubsection{Empty Spaces}
The empty spaces can be presented in two ways: either just a space or a space with variable length. The former is derived using a predefined length threshold, and is suitable for line-delimited parking spaces; while the latter is adaptively obtained by multiplying the time difference of two neighbouring obstacles and the average speed, which is suitable for un-delimited spaces and flexible for compact cars or long lorries.

\subsubsection{Offsetting GPS}
A GPS with high frequency refresh rate is advisable. In order to accurately pinpoint the illegal paring vehicles or empty spaces on the map, the GPS coordinates obtained from the sensor need to be calibrated. In this article, a pre-defined list of outstanding environment signatures is used to offset the GPS drift. In other words, we align the GPS sensor data with the pre-defined environment signatures when the supervised learning algorithm recognizes the environment signatures as the CroPark vehicle drives by.


\section{Empirical Results}\label{sec:evaluation}
In this section, we evaluate the supervised learning algorithm by examining the vehicle and space detection accuracy. In our drive test, the average cruising speed is $20$ km/h, and the updating time is around $120$ seconds, namely, we will pass by the same point every $2$ minutes.
\subsection{Vehicle Detection} 
Table \ref{t1}-(a) gives parked cars' statistics from park1 to park5 with $6$ runs. The data format of $x/y/z$ denotes 'number of detected vehicles/ground truth/parking capacity'. For example, in park1 of the fifth run, there were 3 cars detected by the algorithm, 3 cars actually parked, and 6 parking spaces in total. The last column shows the false positives of each run, which is the number of road clutters detected as vehicles.

The vehicle detection rate is $123/124=99.2\%$, and the accuracy rate is $123/140=87.9\%$ by counting the false positives. Note, the accuracy will increase if the updating frequency is higher or if there are more CroPark units employed, the relationship of which is explored in the following Section \ref{sec:est} and Fig. \ref{Fig:unit}.

\begin{table*}[t!!!!!!!]
\centering
\caption{Drive-test Results}
\begin{tabular}{cc}
    \begin{minipage}{.5\linewidth}
\caption*{(a) Parked Vehicle Detection}
\centering
{\normalsize
\begin{tabular}{c c c c c c c}
\hline\hline
Run no.  &  park1 &  park2 &  park3 &  park4 &  park5 & false positives\\ [1ex]
\hline
1	  	&	$6/6/6$&	$1/1/1$&	$1/1/1$&	$10/10/11$&	$5/5/5$ &$2$\\
2	  	&	$6/6/6$&	$1/1/1$&	$1/1/1$&	$10/10/11$&	$5/5/5$ &$4$ \\
3	  	&	$6/6/6$&	$0/0/1$&	$1/1/1$&	$10/10/11$&	$5/5/5$ &$3$ \\
4	  	&	$2/2/6$&	$0/0/1$&	$1/1/1$&	$10/10/11$&	$5/5/5$ &$3$ \\
5	  	&	$3/3/6$&	$1/1/1$&	$1/1/1$&	$9/10/11$&	$5/5/5$ &$2$ \\
6	  	&	$4/4/6$&	$1/1/1$&	$1/1/1$&	$7/7/11$&	$5/5/5$ &$2$\\
\hline
\end{tabular}
}
\label{t1}
    \end{minipage} &
\\
\\
    \begin{minipage}{.5\linewidth}
\caption*{(b) Space Detection}
\centering
{\normalsize
\begin{tabular}{c c c c c c c}
\hline\hline
Run no.  &  park1 &  park2 &  park3 &  park4 &  park5 \\ [1ex]
\hline
1	  	&	$0/0/0$&		$0/0/0$&		$0/0/0$&		$1/6.86 m/1$&	$0/0/0$\\
2	  	&	$0/0/0$&		$0/0/0$&		$0/0/0$&		$1/6.56 m/1$&	$0/0/0$  \\
3	  	&	$0/0/0$&		$1/7.00 m/1$&	$0/0/0$&		$1/7.23 m/1$&	$0/0/0$  \\
4	  	&	$4/9.39m,9.04m/>1$&	$1/9.62 m/1$&	$0/0/0$&		$1/6.11 m/1$&	$0/0/0$  \\
5	  	&	$3/6.24m,5.54m/>1$&	$0/0/0$&		$0/0/0$&		$2/9.45 m/>1$&	$0/0/0$ \\
6	  	&	$2/6.52m,7.09m/>1$&	$0/0/0$&		$0/0/0$&		$4/7.19 m/>1$&	$0/0/0$ \\
\hline
\end{tabular}
}
\label{t2}
    \end{minipage} 
\end{tabular}
\end{table*}

\subsection{Space Detection} 
\begin{figure*}[t!!!!!!]
\begin{center}
\subfigure[Raw Data]{\includegraphics[scale=0.6]{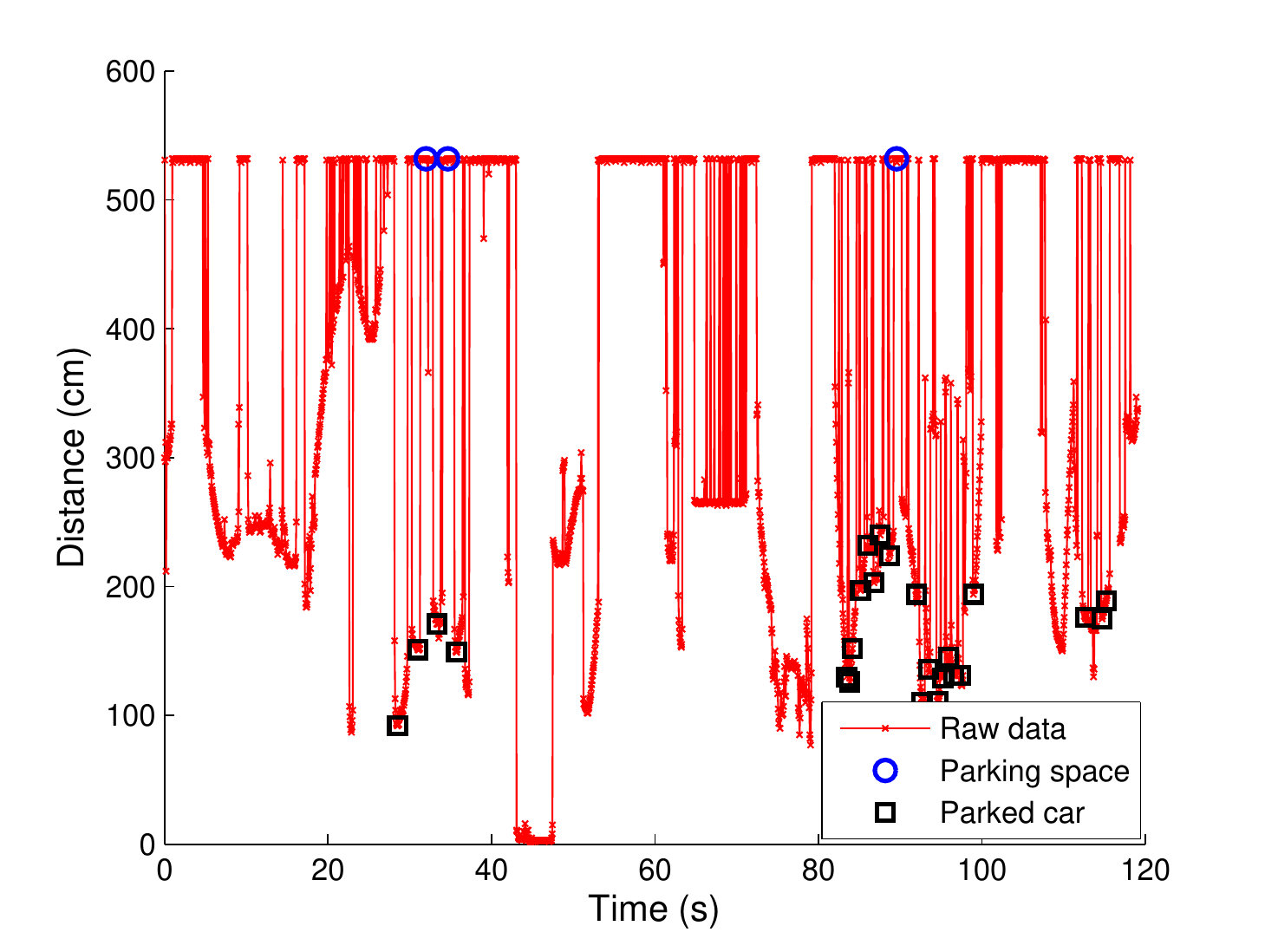}{\label{Fig:ra5}}}
\subfigure[Space Map]{\includegraphics[scale=0.6]{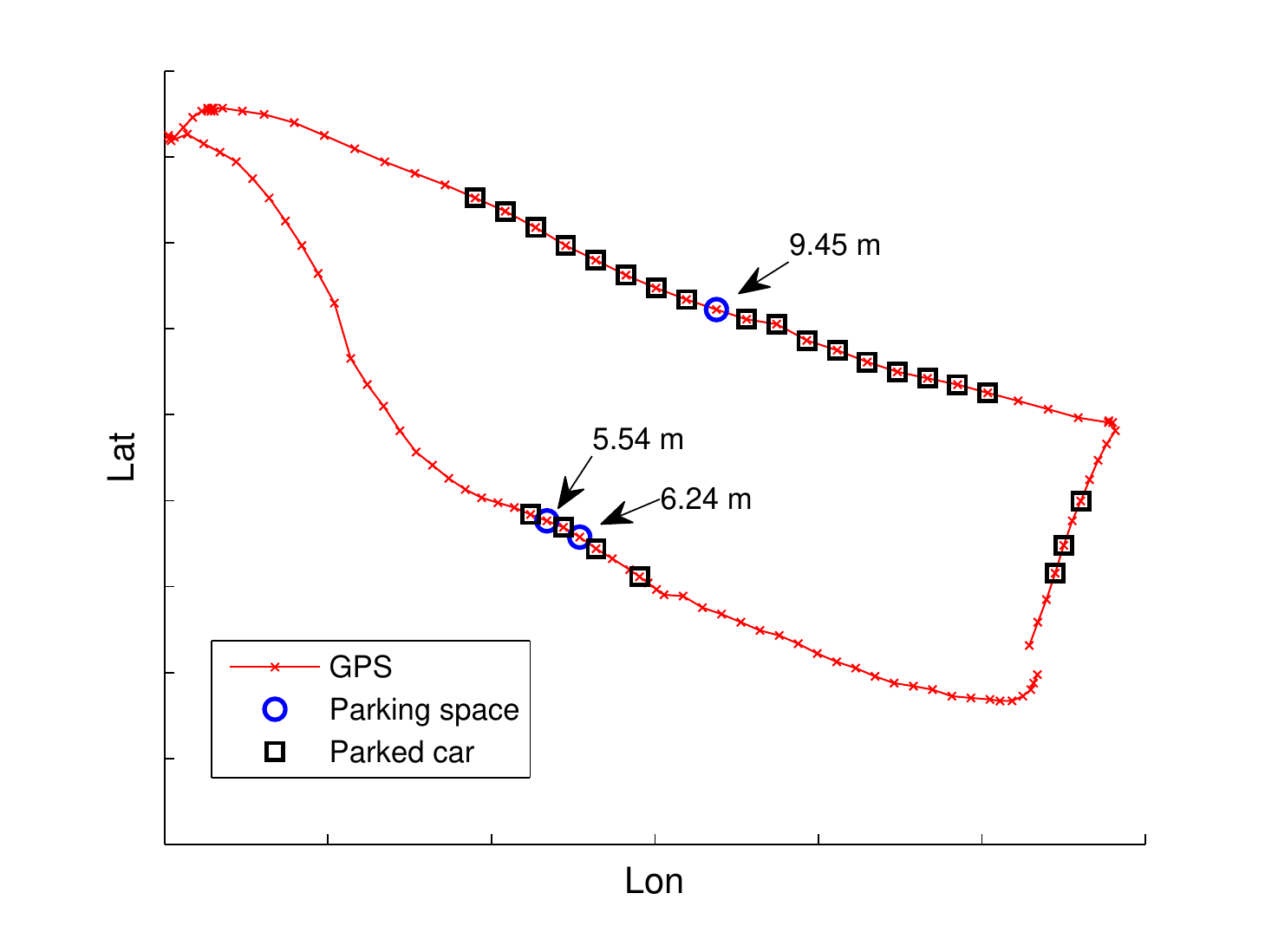}{\label{Fig:sp5}}}
\caption{Trace Plot of Run 5}
\label{Fig:r5}
\end{center}
\end{figure*}

As the on-street parking zones on High Street and Church Road are un-delimited parking, the length and the parking manner of the parked cars will greatly affect the detection results. Therefore, it may not be accurate to infer the number of parking spaces by deducting the number of detected cars from the parking capacity. For this reason, we give parking spaces in two ways, namely, the number of parking spaces by subtracting the number of detected cars from the parking capacity, and parking spaces with detected available length. The data format of $x/y/z$ in the Table \ref{t2}-(b) denotes 'number of detected spaces by subtracting from parking capacity/detected spaces with variable length/ground truth'. For example, in park1 of the fifth run, 3 spaces are inferred by the subtraction, 2 spaces with different length are obtained from the length-auto-adjusting algorithm, and more than 1 space is observed in the recorded video. The trace plot of Run 5 and its space map are shown in Fig.\ref{Fig:r5}. The space map Fig.\ref{Fig:sp5} provides the parking space with extra length information, which is particularly important to customized requests. For example, a Smart for-two Cabrio can fit in a 3-meter space, while a BMW M5 may require a 6-meter space.

\section{Estimating CroPark Units Needed for A City}\label{sec:est}
Let $s$, $p$ and $w$ denote the area of a city, the ratio that total roads account for a city area, and the average road width, respectively. The total road length of the city can be obtained: $L=s\cdot p/w$. If the parking occupancy map updating frequency and the vehicle detection accuracy rate are $\tau$ and $\gamma$, the number of CroPark units $m$ needed  to cover the city can be estimated:  

\begin{center}
\begin{equation}
  m = \frac{L}{\nu\cdot\frac{1}{\tau}\cdot\gamma},
\label{eq1}
\end{equation}
\end{center}
where $\nu$ is the average cruising speed of CroPark vehicles.

In order to compare with fixed sensing techniques, we use this model to estimate the number of CroPark units needed for San Francisco. The upper east part of San Francisco has an area of $19.26 \text{km}^2$, where the SFpark pilot is deployed (excluding the metropolitan area). Assuming urban streets account for around $10\%$ of the city area, and the average street width in San Francisco is $10$ meters. The total road length is then derived as:

\begin{center}
\begin{equation}
\nonumber
  L = \frac{19.26\cdot10^6\cdot10\%}{10}=1.926\cdot10^5 \ \text{meters}.
\label{eq2}
\end{equation}
\end{center}

If the parking occupancy map is updated every $2$ minutes (i.e., $\tau=0.0083$), the vehicle detection accuracy rate is $\gamma=87.9\%$, and the average city cruising speed is assumed to be $\nu=20\ \text{kilo meters per hour}$ (i.e., $5.55$ meters per second), the required number of CroPark units to be mounted on vehicles is:
 \begin{center}
\begin{equation}
\nonumber
  m = \frac{1.926\cdot10^5}{5.55\cdot120\cdot87.9\%}=328,
\label{eq1}
\end{equation}
\end{center}
\begin{figure}[h!!!!!!!!]
\centering
\includegraphics[scale=0.62]{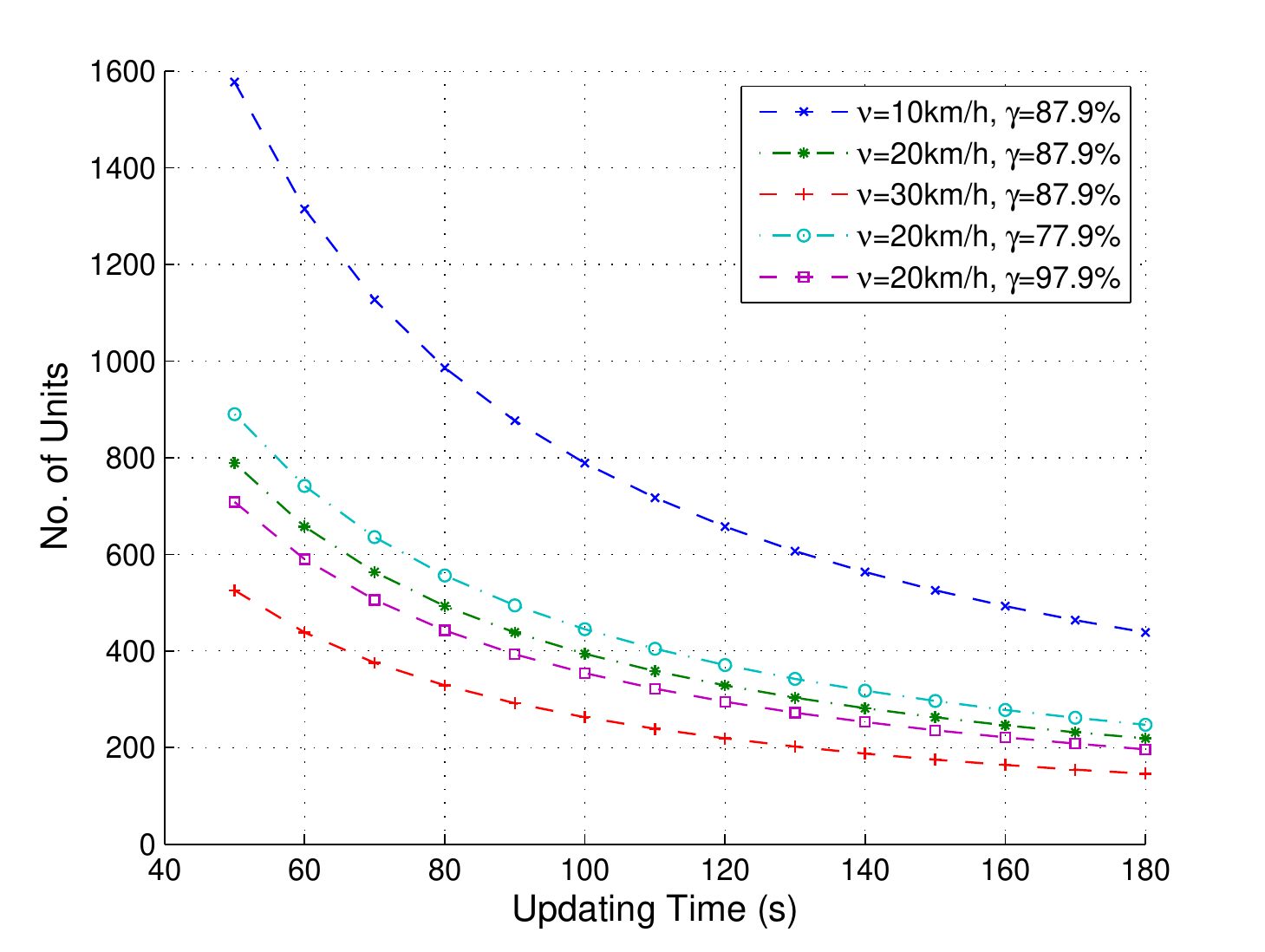}
\caption{Number of Units against Updating Time}
\label{Fig:unit}
\end{figure}
which is significantly lower than $12000$ sensors deployed for $8000$ parking spaces in the SFpark pilot area. 

Note that, this simple estimation model is indifferent to the number or spatial distribution of parking bays, as all points of the roads will be covered by a $\nu\cdot\frac{1}{\tau}$ rolling window with frequency $\tau$ and accuracy $\gamma$. 


Fig.\ref{Fig:unit} also illustrates how many CroPark units are needed against updating time (i.e., $1/\tau$) as cruising speed $\nu$ and the vehicle detection accuracy $\gamma$ vary.

\section{Conclusions}\label{sec:conclusions}
In this article, we present CroPark that is consisted of an ultrasonic ranger and a GPS receiver to determine the distance from the vehicle to roadside. It is proposed that CroPark units could be placed on vehicles such as buses, taxis or private cars to continually gather data as they travel along their routes. CroPark leverages the users' heuristic searching and converts users to stakeholders. On the one hand, CroPark requires much lower number of sensor units compared with the fixed sensing solutions thanks to users' participation; on the other hand, CroPark cuts down the time and fuel consumed in searching thanks to the dissemination of parking availability information. As soon as the initial occupancy map is built up in the crowdsourcing manner, CroPark can substantially bring down the searching effort.

As for next steps, lane-changing detection, a balanced guidance algorithm and a right price model for on-street parking need to be considered.
\begin{enumerate}
\item{\textbf{Lane-changing detection}}: The designed algorithm and current sonar configurations can only handle the single lane situation. If the vehicle moves to a second lane, then the sonar may misunderstand the extra distance as empty parking spaces. One solution is to employ a high-precision GPS, however, it may be difficult as the adjacent lanes are within just a few meters. The other solution could be employing dual ranger sensors positioned at each end of the car to measure the differentials. For example, if the vehicle moves away from one lane to another, then the difference of the two sensors' reading remains constant if we adjust the lateral time difference.
\item{\textbf{Balanced guidance algorithm}}: Owing to the updating delay and some parking areas are more preferable than others, people may flock to the same places and then find the spaces already occupied. One solution is to provide future occupancy prediction based on open data or history data \cite{7361177}. The other solution is to design a reservation system or a balancing algorithm that guilds people to different locations if multiple requests to the same space are received. 
\item{\textbf{Right price model}}: As pointed out in \cite{shoup2006cruising}, drivers are actually encouraged to circle blocks to find on-street parking spaces if the kerb parking is free or under-priced. In other words, drivers flock to the cheap kerb parking spaces, and thus create more cruising and congestions. This suggests further research is needed to find the right price model for on-street parking. 
\end{enumerate}



\bibliographystyle{ieeetr}	
\bibliography{ref}

\end{document}